\documentclass[fleqn]{jleo}


\usepackage{amssymb,graphicx,amsfonts,bm,graphicx}
\usepackage{times,amsmath,tnqupgreek}

\usepackage{threeparttable}
\usepackage{multirow}
\usepackage{array} 
\usepackage{longtable}

\usepackage{booktabs} 

\usepackage{hyperref}
\begin{document}

\title{Fintech and MSEs Innovation: an Empirical Analysis}

\author{Siyu Chen, Qing Guo\setcounter{footnote}{1}\thanks{$\langle${qingguo@ruc.edu.cn}$\rangle$
\newline \hspace*{10.5pt}We are grateful for the advice from the Conference on Fintech, Innovation and Development (CFID) 2021, the dataset of China Micro and Small Enterprise Survey (CMES) from Southwest University of Finance and Economics, and the Index of Digital Financial Inclusion from Peking University. 
}\\
Renmin University of China,\\ Beijing, China}

\maketitle


\pagestyle{headings}
\begin{abstract} 
{\textbf{Abstract:} Employing a comprehensive survey of micro and small enterprises (MSEs) and the Digital Financial Inclusion Index in China, this study investigates the influence of fintech on MSE innovation empirically. Our findings indicate that fintech advancement substantially enhances the likelihood of MSEs engaging in innovative endeavors and boosts both the investment and outcomes of their innovation processes. The underlying mechanisms are attributed to fintech's role in fostering long-term strategic incentives and investment in human capital. This includes the use of promotions and stock options as rewards, rather than traditional perks like gifts or trips, the attraction of a greater number of university graduates, and the increase in both training expenses and the remuneration of technical staff. Our heterogeneity analysis reveals that fintech exerts a more pronounced effect on MSEs situated in economically developed areas, those that are five years old or younger, and businesses with limited assets and workforce. Additionally, we uncover that fintech stimulates the innovation of MSEs' independent research and development (R\&D) efforts. This paper contributes to the understanding of the nuanced ways in which fintech impacts MSE innovation and offers policy insights aimed at unleashing the full potential of MSEs' innovative capabilities.
}
{\textbf{Key words:} Fintech; innovation; micro and small enterprises (MSEs); R\&D; strategic incentives; investment to human capital }
\end{abstract}


\section{Introduction}

Micro and small enterprises (MSEs) are key players in the business field, as they make a significant contribution to the innovation, output and employment. According to the ILO data from 99 countries, 70\% of employment comes from MSEs and the self-employed. MSEs play an irreplaceable role in promoting economic growth.

Technology innovation is the core competitiveness of MSEs for the increasingly fierce international competition, and it is also an important internal driving force for a country’s high-quality economic growth. 

The digital finance ecosystem is undergoing a metamorphosis, largely propelled by the integration of machine learning technologies. Machine learning, with its ability to discern intricate patterns from vast datasets, has become a cornerstone for enhancing the efficiency, accuracy of financial services\cite{xin2024vmt, chen2021deep, zhou2021modeling, zhou2024optimizing, 
xiao2024simple}.

Machine learning models can analyze vast amounts of data to predict and assess credit risks more accurately. This allows financial institutions to make informed lending decisions, particularly for MSEs that may lack extensive credit histories. By leveraging customer data, machine learning algorithms can identify individual preferences and behaviors, enabling financial institutions to offer personalized products and services tailored to the unique needs of each customer\cite{zhang2024deepgi, chen2021multi, zhou2022eventbert}.

However, the financial markets of most developing countries are still in the primary developing stage for the lack of collateral, complicated procedures, credit rent-seeking, etc., which make it difficult for MSEs to obtain financial funds and may hinder their innovation vitality. In addition, it further aggravates the financing difficulties of MSEs that the traditional credit financing favors state-owned and larger enterprises. According to the International Finance Corporation (IFC), 40\% of formal micro, small and medium-sized enterprises (MSMEs) in developing countries have \$5,200 billion of unmet financing needs each year. East Asia and the Pacific region accounts for the largest share of the global finance gap (46\%), which is followed by Latin America and the Caribbean (23\%), and Europe and Central Asia (15\%).

Hence, it is important to make up for the shortcomings of traditional financial services and release the research and development (R\&D) vitality of MSEs under the background of global innovation-driven development, especially for developing countries and emerging economies. In the context of the global Corona Virus Disease 2019 (COVID-19) pandemic and the urgent need for economic recovery, the importance of activating the vitality of MSEs as a key point to stimulate the economy is self-evident.


With the arrival of the era of big data, cloud computing and artificial intelligence, digital technology is widely used in many countries around the world\cite{zhang2024development, xiao2021learning, chen2024pareto, xu2024research, zhou2021improving, zhou2024predict}. The development of digital technology promotes MSEs service innovation and data innovation\cite{chen2021pareto, xiao2022decoupled, zhou2024visual, chen2022ba}.
Alipay and Ant Credit Pay, the world's leading open financial technology platform from Ant Financial Services Group, are typical cases of the application of digital technology in the financial field. Relying on digital technology, Alipay and Ant Credit Pay provides inclusive financial services for MSEs and individual consumers, which greatly promote the development of inclusive finance in China, and become an important practice for China and even the world to promote inclusive finance. At the end of the year 2015, the State Council of China issued the Notice of Promoting the Development Plan for Inclusive Finance (2016–2020), and made the development of inclusive finance a national strategic plan, which lead to a rapid development of fintech in China. According to the Digital Financial Inclusion Index of China (PKU-DFIIC) calculated by the Digital Finance Research Center of Peking University\cite{guo2020measuring}, the average PKU-DFIIC of each province rose from 40.004 in 2011 to 300.208 in 2018. Compared with traditional finance, fintech has better geographical penetration and stronger user accessibility, which can effectively reduce the financing costs of MSEs and expand the coverage breadth of financial services. The emergence and rapid development of fintech in China injects a strong impetus to the innovation and development of MSEs, and provides a good reference for other countries in the world to stimulate innovation and economic recovery.

The rapid development of fintech begins to attract academic attention. 
The evolution of information and communication technology, big data, the Internet of Things, and cloud computing has ushered in a new era of business paradigms, significantly boosting the operational efficiency within the commercial sector.

The contributions of this paper lie in the following aspects. Firstly, we focus on the MSEs innovation instead of listed companies and verifies the inclusive impact of fintech on MSEs. The existing study pay attention to the impact of fintech on the listed companies, but the listed companies can use formal credit channels to get the loans, and few listed companies use the online channels for financing. The main beneficiaries of fintech are MSEs but by now there is no research focusing on the impact of fintech on MSEs. This paper focuses on the impact of fintech on MSEs innovation. Secondly, This paper further studies the influence mechanisms from the enterprise internal perspectives and the aspects of personnel economics. We explore how the fintech promote the long-term strategic incentives and investment to human capital, including promotion and stock as the incentives, recruiting university graduate students, increasing training fee and the incomes of technicians, so as to promote MSEs innovation. Thirdly, for the theoretical contributions, this paper expands the research on financial development and enterprise innovation from the perspective of fintech. Some of previous studies have examined the effect of traditional finance on enterprise innovation. This paper examines the impact of the integration of emerging technology and traditional finance on enterprise innovation from the perspective of fintech, which is an important supplement to previous studies from the theoretical aspects. Fourthly, The China Micro and Small Enterprise Survey (CMES) data which focuses specifically on MSEs makes it possible for us to answer the above questions. This paper systematically investigates the impact of fintech development on the innovation of MSEs, and further analyzes the mechanisms of fintech promoting innovation of MSEs by combining the PKU-DFIIC and CMES at the micro level. Under the situation of the COVID-19 pandemic, MSEs should be released to innovate and stimulate the market economy, so as to provide a source of vitality for global economic recovery and high-quality development.

We choose China--the largest developing country as the research object to examine the impact of fintech on the innovation of MSEs based on following three key reasons. First, fintech has developed rapidly in China in recent years. Ant Financial take its total above 500 million clients, almost 10 times the level of the world's biggest banks (Arnold 2017). China's digital financial development occupies a leading position in the world in terms of industry financing scale and application scenarios (Tang, Wu, and Zhu 2020). Second, China has a more inclusive regulatory environment for the fintech. Since 2015, when the development of inclusive finance was included in the national strategy, the Chinese government has successively introduced a series of specific measures to encourage the development of fintech. Technology companies have entered the financial service field on a large scale, and a lot of digital financial institutions have emerged. Third, China has a large scale of MSEs, and by the end of 2017, China's MSEs had reached 73.281 million (Wang 2019), but the traditional financial market in China is underdeveloped, which makes many MSEs be excluded from the formal credit. It leaves a vast unfilled market for fintech (Saal, Starnes, and Rehermann 2017). These above reasons make China an suitable and excellent research object to study whether and how fintech promotes innovation of MSEs.

The following arrangement is as follows. Section 2 is the literature review. Section 3 is the theoretical analysis. Section 4 expounds the data source, the main variable selection and model setting, and section 5 reports the empirical results and explores the possible impact mechanisms, and discusses the heterogeneity influence. Section 6 finally draws a conclusion and summarizes the full paper.

\section{Literature review}



The digital finance landscape has been profoundly reshaped by the advent of machine learning, a subset of artificial intelligence that empowers financial institutions to process vast amounts of data with remarkable efficiency and accuracy\cite{chen2022ba, zhou2022claret, zhou2023towards, wang2020global}. This technology has not only revolutionized the way financial services are delivered but has also become a catalyst for development in the MSE sector\cite{liu2024infrared}.

Machine learning algorithms have transformed credit scoring and risk assessment models, making them more accurate and inclusive. By analyzing patterns and trends in data, these models can predict the creditworthiness of MSEs with greater precision, thereby enhancing their access to capital. This has been particularly beneficial for MSEs that traditionally struggled to secure loans due to a lack of credit history\cite{jiang2022role, chen2021improving, chen2022ba, chen2021pareto}.

The ability of machine learning to analyze individual customer data has led to the development of personalized financial products and services. MSEs can now access tailored financial solutions that cater to their unique needs and circumstances, which not only improves customer satisfaction but also strengthens the competitive edge of these enterprises.

Machine learning techniques are crucial for addressing the inherent data scarcity challenges faced by MSEs\cite{lv2023ideal, chen2024learning, xin2024mmap, xin2024parameter, xin2023self, zhang2024enhancing}. By employing algorithms that can generalize well from a small set of examples, these enterprises can leverage the power of machine learning to make informed decisions, optimize operations, and enhance their financial services\cite{xu2024research, chen2021pareto, zhang2024development, liu2024enhanced}.


The impact of fintech on economic development has attracted the attention of scholars in recent years. The current research mainly focuses on the impact of fintech on entrepreneurship, residents' consumption, urban-rural income gap and inclusive growth. At the micro level, digital finance narrows the gender earnings gap\cite{2021Does}. Digital finance can promote individual consumption by easing liquidity constraints\cite{gazel2021entrepreneurial}. At the same time, fintech promotes the household consumption\cite{hau2017techfin}, and narrows the urban-rural income gap\cite{zhang2020trickle}.


From the perspective of generalized economic factors, Schumpeter, as one of the earliest economists who focuses on enterprise innovation, investigate the influence of enterprise scale and market structure on enterprise innovation\cite{1942Capitalism}. Later, it is introduced that factors such as enterprise characteristics including enterprise human capital. Barney (1991)\cite{1991Firm} hold the opinion that valuable and scarce resources, such as human capital\cite{H2002OVERCOMING}, technology or skills,
 can drive enterprise innovation. Some scholars study the influence of social network on enterprise innovation\cite{2020Enterprise}.

According to existing research, it can be found that, firstly, there have been a wide range of influencing factors on enterprise innovation, and the existing research mainly includes economic and financial factors, business management factors, policies and institutional environment, as well as culture and social networks and other factors. However, from the perspective of the digital economy, research on the promotion of enterprise innovation by the development of fintech is still rare and needs to be further studied. Secondly, at present, there are a few of research related to this study, and all of them take listed companies as research objects. Because of the companies applying to list need to have a certain scale, while the focus of inclusion financial policy is to serve MSEs, the impact of fintech development on MSEs and their transmission mechanisms should be highlighted. We study MSEs to explore whether fintech can play an inclusion role in helping MSEs to innovate, and to further analyze and explore their influence mechanisms.

\section{Theoretical analysis and research hypothesis}

\subsection{Fintech and MSEs innovation}

Theoretically, according to Schumpeter's innovation theory, enterprise innovation input cannot be separated from certain social and economic conditions and needs the support of effective financing system\cite{1942Capitalism}. Enterprises engaged in R\&D and innovation activities need to be financed through bank credit. However, due to various deficiencies in traditional bank credit, such as information asymmetry, banks cannot obtain all the information of enterprises. For risk avoidance, banks prefer to lend to large enterprises, which may lead to severe financing constraints of MSEs. Fintech can provide MSEs with detailed transaction records as credit endorsement, thus reducing the cost of bank identification. Furthermore, relying on the popularity of digital technology, fintech improves the user's access rate and enables MSEs owners to obtain financing services more conveniently, thus reducing financing constraints. The support of fintech enables MSEs to obtain the funds required for innovation activities so as to promote their innovation activities.

Machine learning algorithms can process and analyze large datasets to uncover insights that might otherwise be hidden. These insights can inform business strategies, identify new market opportunities, and guide product development\cite{chen2024pareto, zhou2021improving, lai2023detect, li2024feature, li2023stock}.

By analyzing customer interactions and transaction data, machine learning can help enterprises understand customer preferences and behaviors. This understanding can lead to the creation of innovative products and services that meet unmet customer needs.

Machine learning can segment customers into different groups based on their behavior, preferences, and demographics. This allows for targeted marketing campaigns that are more effective and innovative, enhancing customer engagement and loyalty\cite{zhou2021modeling}.

Automation of processes through machine learning can reduce costs and increase efficiency, freeing up resources for enterprises to invest in innovative projects and research and development.


Based on this, we put forward the following research hypotheses:

H1: Fintech can promote the participation of MSEs in R\&D and innovation activities and increase the innovation input and output.

\subsection{Fintech and MSEs strategic incentives and recruitment to human capital}

Enterprise innovation need stable human capital input. Investment in innovation requires financial resources\cite{2011R}. Strong financing constraints will inhibit enterprise innovation. While the emergence of fintech greatly reduces the financing costs and improves the availability of financing for enterprises, thus greatly reduces the financing constraints faced by enterprises\cite{2021Does}, and provides stable financial support for enterprise R\&D and innovation activities. Sufficient capital investment help enterprise improve the incentive strategy and adjust the staffing structure.

Enterprise human capital is a valuable and scarce resource, which plays the key role in enterprise innovation.
Motivating human capital in knowledge-intensive activities is a serious managerial challenge because it is difficult to link rewards to actions or performance. Firms instead might motivate knowledge workers by offering them opportunities to increase personal benefits through autonomy in the decision-making process\cite{2015Strategic}. When the proportion of professional technicians increases and the incentive of professional technicians enhances, enterprises will gain strong innovation power. Staff training will also promote human capital accumulation to achieve enterprise innovation. Since professional technicians are highly skilled workers and their salaries are a major component of R\&D spending\cite{2011R}. Hence, MSEs have to pay high labor costs for innovation activities. As mentioned above, under the traditional financial background, MSEs are facing strong financing constraints. When MSEs are faced with expensive technical personnel input, it is difficult to obtain effective financing support, which leads to the lack of human capital investment of MSEs. Fintech has greatly reduced the threshold for MSEs to obtain external financing, thus enabling them to invest effective human capital to support innovation activities.

As machine learning becomes more integrated into digital finance, there is a growing demand for employees with skills in data analysis, machine learning algorithms, and artificial intelligence. This drives the development of new training programs and educational opportunities for employees\cite{chen2021pareto, xiao2022decoupled, chen2022ba}.
Automation of routine tasks through machine learning can free up employees from time-consuming, manual processes, allowing them to focus on more strategic and creative tasks that require human expertise and judgment\cite{lv2023duet, lai2024selective, zhou2024visual, liu2024enhanced, yanhui2024dog}.

The increased use of machine learning in decision-making processes necessitates a higher level of data literacy among employees. This includes understanding how to interpret data, analyze results, and make informed decisions based on insights derived from machine learning models.

Machine learning can provide employees with advanced analytics and predictive insights, which can improve decision-making capabilities across various levels of the organization, from management to frontline staff. By analyzing the unique economic and social conditions of a region, machine learning can help design financial products tailored to the specific needs of MSEs in these areas. Machine learning can analyze market trends and consumer behavior to help MSEs identify opportunities for innovation and growth within their local economies. Automation of financial processes through machine learning can reduce costs for MSEs, allowing them to allocate more resources towards innovation and growth\cite{gai2021multi, chen2023mapo, zhou2022eventbert, Shen2024Harnessing}.

Based on this, we put forward the following research hypotheses:

H2a: Fintech promotes the innovation of MSEs by promoting long-term strategic incentives to MSEs human capital.

H2b: Fintech promotes the innovation of MSEs by promoting strategic recruitment to MSEs human capital.

\subsection{Fintech and different types of MSEs innovation}

Due to the information asymmetry of financing, traditional finance prefers to lend to elderly and larger enterprises, leading to financing discrimination for younger and smaller enterprises. At the same time, MSEs are difficult to obtain bank credit due to lack of collateral. As a result, enterprises with fewer assets and employees suffer more from financing constraints than larger companies.
Machine learning can help MSEs protect their financial transactions from fraud, ensuring the security of their business operations\cite{xiao2021learning, chen2024pareto, zhou2022claret, yu2024credit, zheng2024advanced, cao2024rough}. Machine learning can assist MSEs in optimizing their supply chains, reducing costs, and improving the efficiency of their operations, which can be particularly beneficial for for enterprises with fewer employees. Machine learning can provide MSEs with insights into customer preferences and behavior, enabling them to develop innovative products and services that cater to local needs\cite{chen2019deep, jiang2022role}.

The emergence of fintech makes it possible to record every transaction in detail, which will reduce the cost of risk identification caused by information asymmetry. At the meantime, fintech has stronger accessibility, which improves the financing availability of younger and smaller enterprises. The characteristics of fintech, such as wide coverage, low threshold, and strong flexibility, are more in line with the financing needs of young and small scale enterprises. Therefore, younger and smaller enterprises excluded from traditional finance could rely more on the financing method of fintech, and the innovation effect of fintech on these enterprises will be more obvious. Machine learning algorithms can analyze limited data sets to assess creditworthiness, allowing MSEs in less developed areas to gain access to loans and credit facilities that might otherwise be inaccessible\cite{zhou2024optimizing, 
unknown1, wang2024graph, article}. Digital finance platforms, powered by machine learning, can extend their reach to remote areas, providing financial services to populations that were previously unserved or underserved. Machine learning models can help in assessing the risk profiles of MSEs in underdeveloped regions, enabling financial institutions to make more informed lending decisions and support viable business ventures\cite{xiao2023reconsidering, chen2021improving, zhou2023towards}.

Online platforms that use machine learning to customize educational content can help MSEs in less developed areas upskill and access training in digital finance and related fields. Machine learning can be used to develop tools that enhance digital literacy, helping MSEs in underdeveloped regions to navigate and benefit from digital financial services. Machine learning can help MSEs in less developed areas to understand and comply with financial regulations, reducing the risk of penalties and legal issues\cite{xiao2022decoupled, xiao2022representation, chen2023invariant, huang2024research}.

Besides, the popularization of fintech depends on the spread of digital technology. MSEs in cities with a wide range of digital technology and a deep degree of use are more likely to use fintech. And for those MSEs located in relatively disadvantaged areas of digital technology, the probability of using fintech to obtain effective financing is lower. Hence, the role of fintech in innovation of enterprises located in areas with developed digital technology will be relatively more obvious. Fintech platforms can use machine learning to connect MSEs with potential partners, investors, and customers, fostering collaboration and innovation. Machine learning can help identify investment opportunities in less developed areas, attracting capital to MSEs that demonstrate potential for growth and innovation\cite{
zhou2024application, Weng2024, zhu2021twitter}
Machine learning can support MSEs in developing sustainable business models that address social and environmental challenges, aligning with global development goals\cite{gai2019deep, chen2021adaptive, Weng202404, li2023scigraphqa}.

Based on this, we put forward the following research hypotheses:

H3a: The promotion effect of fintech on the innovation of MSEs is more significant in the regions with developed digital technology.

H3b: The promotion effect of fintech on innovation of MSEs is more significant for enterprises with shorter established years.

H3c: The promotion of fintech on innovation of MSEs is more significant for enterprises with fewer assets.

H3d: The promotion of fintech on innovation of MSEs is more significant for enterprises with fewer employees.

\section{Data Source, Variable Selection and Model Setting}

\subsection{Data Source}

The data used in this paper are as follows: (1) The Peking University Digital Financial Inclusion Index of China (PKU-DFIIC). This index is produced by a research team from the Institute of Digital Finance at Peking University and Ant Financial Services Group (the index appendix is shown in Appendix 1). The index is compiled by micro data on fintech from Ant Financial Services Group, China's representative internet financial institution, and aims to provide a scientific and accurate picture of the development of fintech in China. It includes provincial and city-level indexes  including the total index of fintech, as well as three sub-indexes: coverage breadth, usage depth and digitization level. Among them, usage depth involves sub-indexes such as payment, credit, insurance, credit investigation, investment, and money funds. (2) China Micro and Small Enterprise Survey (CMES) database. This data is provided by the Survey and Research Center for China Household Finance of Southwest University of Finance and Economics, which conducted a nationwide large-scale sample survey of MSEs in 2015, including small enterprises with independent legal personality, micro enterprises and family-owned enterprises. The survey obtained a total of more than 5400 MSEs covering 28 provinces (autonomous regions and municipalities directly under the Central Government) in China, with national representation. (3) The regional economic development index of this paper comes from the China Statistical Yearbook. After data processing, we finally obtain 1533 observations of MSEs.

\subsection{Variable Selection}

The dependent variable for this paper -- "innovation of MSEs" is from the database of CMES 2015, which uses the question of "whether there are products or technology research and development and innovation activities" to objectively measure and 1 and 0 represent yes and no respectively. To identify the subsequent and timely impact of fintech on MSEs innovation, we use the index of MSEs innovation activity participation to measure MSEs innovation. Considering the long application and approval cycle of patents, it is not conducive to use the index of patent number to identify the timely response of the impact of fintech on MSEs. As for the independent variable "fintech", this paper selects the provincial-level index (2014), which lagged one period behind China's digital inclusion financial index, as the core independent variable to investigate the influence of fintech index on the innovation of MSEs. This is considering that the development of fintech has a certain lagged effect on the enterprises and it can also reduce the endogeneity problems caused by reverse causality. This paper also selects two sub-indicators, fintech coverage breadth and usage depth, and replace the independent variable with the fintech index with a lag of 2 years (2013) and the survey year (2015), to further examine the robustness of the fintech’s impact on MSEs innovation.

In addition, according to the data from the China Micro and Small Enterprise Survey (CMES) and the practice of existing research, other variables that may influence enterprise innovation have been controlled in this study. Specifically, two categories of variables were mainly controlled: enterprise characteristics and regional characteristics. Enterprise[ We use enterprise to indicated micro and small enterprise in the full text.] characteristics include the logarithm of the enterprise total assets, enterprise age, enterprise size, whether the enterprise is a high-tech enterprise, government subsidies, the age of the enterprise owner, the education years of the enterprise owner, the gender of the enterprise owner, enterprise ownership, and the industry the enterprise belonging to. Regional characteristics include the level of macroeconomic development measured by provincial GDP, and the regional effect. Additionally, the provincial population size by the end of the year, provincial human capital level, provincial science and technology public expenditure were used to run the robustness test.

Specific variable definitions and descriptive statistics of the main variables are shown in Table 1 and Table 2.

\subsection{Fintech and MSEs innovation}

We use linear probability model (LPM), ordinary least squares (OLS) and the Probit model to conduct analysis. For the regression of innovation activities, successful patent application, we use the LPM, and for the regression of R\&D expenses and increasing revenue from R\&D, we use the OLS model. 
    (1)
In model (1), i represents the enterprise identity document (ID), j represents the province where the enterprise is located in, and t represents the year. innovation refers to the dependent variable, that is "whether the enterprise has successfully applies patent". index refers to the independent variable, the fintech index that lagged one year (year of 2014). CV is the series of control variables all above at enterprise level and regional level. own represents the effect of enterprise ownership. ind represents the effect of industry. region represents the effect of region. Among them, the control variables at the regional level are consistent with the fintech index and the data are lagged one year (year of 2014). By this way, we investigate the impact of the development level of fintech in the previous period on the innovation of MSEs in the current period.
 In the robustness test, since the dependent variable "patent application" in the robustness test is a dummy variable, the following probit model is constructed for empirical estimation:
 (2)
In model (2), i represents the enterprise identity document (ID), j represents the province where the enterprise is located in, and t represents the year. innovation refers to the dependent variable, that is "patent application" and "product innovation". The meaning of other variables is the same as model (1).

\section{Empirical Analysis and Discussion}

\subsection{Results of benchmark regression}

We investigate the impact of fintech development on innovation of MSEs according to the LPM and OLS model. Table 3 reports the benchmark regression results of fintech index on enterprise R\&D and innovation input and output, in which column (1)-column (4) show the regression results of innovation activities, successful patent application, R\&D expenses and the increasing revenue from R\&D respectively. It can be seen from the estimation results that the higher the fintech index is, the higher the probability and level of MSEs participating in R\&D as well as the innovation input and output is, and this effect is significant at the 1\% statistical level. This indicates that the MSEs participating in R\&D and the innovation input and output increase with the development of fintech. In other words, the development of fintech can indeed significantly promote the participation of MSEs in R\&D and innovation activities. The results support Hypothesis 1.

From the perspective of other control variables, the larger the scale of the enterprise are, the higher probability and level of engaging in R\&D and innovation activities is. High-tech enterprises as well as government-subsidized enterprises are also more likely to engage in R\&D activities. At the same time, we find that the personal characteristics of enterprise owner also have an impact on innovation. For example, enterprises with a longer-educational-year owner and with a male owner are more likely to engage in R\&D.

\subsection{Robustness test}

In the above, we use the method of Probit and OLS estimation models to test the robustness of the benchmark regression results using the indicators of patent application, independent R\&D expenses and the product innovation. Since the estimated coefficient of direct regression of the Probit model is not of economic significance, the marginal effect is calculated and reported in this paper. It can be seen from the estimation results that the higher the fintech index is, the higher the probability of MSEs product innovation and patent application is, and the independent R\&D expenses is increasing with the fintech development. The effects are significant at the 1\%-10\% statistical level.

To avoid the sample self-selection bias, we use the sample of MSEs that established more than 5 years to conduct robustness test. Considering outstanding enterprises with the potential for success may need the support of financial technology to decrease the financial difficulties, so these enterprises may be inclined to choose regions with higher financial technology level to carry out R\&D activities. According to the establishment year of the MSEs, we choose the MSEs which established before fintech development (the MSEs age over 5 years) as the analysis object. And we find that most of enterprises were established before the development of regional digital finance, and the sampling strategy of CMES data is intra-regional random. The results further prove the robustness of the conclusions.

Next, we use the method of replacing independent variables to further test the robustness of the results. That is to replace the fintech index (2014) in benchmark regression with the coverage breadth, usage depth, and the lagging 2 years (2013) and the survey period (2015) fintech index. The results are shown in columns (1)–(4) of Table 6. As can be seen from the Table 6, when the independent variable is replaced with the above-mentioned different measurement indicators, the development of fintech can still improve the probability of innovation of MSEs and this effect is significant at the 1\% statistical level, which further proves the robustness of the conclusions.

In the benchmark regression, although we have controlled the factors of enterprise characteristics that may affect the innovation of MSEs as much as possible, considering that there may still be some missing factors at the regional level, we further take more regional level factors into account. Based on the benchmark regression model, we further add the provincial population size by the end of the year, provincial human capital level measured by the average number of students in higher education schools per 100,000 people in one province, and the provincial public expenditure on science and technology as control variables. As shown in the column (5) of Table 6, after controlling more regional factors, the development of fintech still significantly promotes the innovation of MSEs, which indicates that the conclusions of our study is robust and reliable, and support Hypothesis 1.

\subsection{Endogeneity test}

As mentioned above, we adopt the practice of lagging the fintech index for one period in empirical study, which alleviates the endogeneity problem caused by the reverse causality, and we also try to control more variables at the regional level to alleviate the endogeneity problems caused by the possible missing variables. But there may still be some unobservable factors that lead to biased estimation results. In order to further solve the endogeneity problems that may exist in this study, the method of Two Stage Least Squares (2SLS) is used in the following part of this paper.

There are two instrumental variables (IVs) used in this paper: one is the coastal port, and the other is the regional internet penetration rate. First of all, The closer to the coastal port, the more complete the industrial organization structure and the higher the degree of economic development of the area, which is more conducive to incubating and activating the new industrial organization form of the digital economy.  Therefore, the distance to the coastal port is related to the development level of fintech in this region, which meets the correlation condition of IV. Secondly, the distance to the coastal port has no direct impact on whether MSEs engage in R\&D and innovation activities, so it meets the exogenous conditions. In addition, referring to the existing research, we use the internet penetration rate as the IV of the development degrees of fintech to conduct further analysis, and the two-stage least square method (2SLS) is used in the analysis.

Table 7 reports the 2SLS estimation results using the distance to port and the internet penetration rate as instrumental variables. Column (1) is the estimation result of the first stage. It can be seen from Table 7 that the coefficient of "distance to coastal port" is significantly negative at the 1\% statistical level, indicating that the further distance of MSEs’ location to coastal port, the lower the level of fintech development. The estimated coefficient of "internet penetration rate" is significantly positive at the 1\% statistical level, indicating that the higher internet penetration rate, the higher development level of fintech. Column (2)-(5) of the Table 7 reports the estimation results of 2SLS, which shows that the impact of fintech on the innovation of MSEs is still significantly positive at the 1\%-10\% statistical level after solving potential endogeneity problems using the method of IV. And the Cragg-Donald Wald F statistic and the Sargan statistic show the effectiveness of the IV method. The results are robust after dealing with the problem of potential endogeneity and support Hypothesis 1.

\subsection{Mechanism analysis}

According to the model (1) and model (2) mentioned above, we further investigate the impact mechanisms of the development of fintech on the innovation of MSEs to test hypotheses H2a and H2b. Table 8 and Table 9 respectively report the estimation results of the two impact mechanisms tested by fintech on strategic long-term incentives to MSEs human capital upgrading and strategic Investment to MSEs human capital.

\subsubsection{Strategic long-term incentives to MSEs human capital upgrading}

The development of fintech can directly help MSEs reduce financing constraints and adjust MSEs incentive strategy of employees. The estimation results are reported in Table 8. The explained variables in columns (1)–(4) of Table 8 are successively the incentive strategy of employees including income, stock, promotion and real rewards. It can be seen from the results that the estimation coefficients of the fintech index are positive in columns (1)–(3), but the coefficient is negative in columns (4), indicating that the development of fintech can significantly promote the adjustment of inventive strategy. Specifically, fintech development increase the probability of MSEs choosing the long-term incentive strategy of employees--such as promotion, stock and income instead of choosing short-term incentive strategy--such as real rewards including gifts and trips. The incentive strategy adjustment may increase employees long-term personal benefits to link rewards and personal goals to long-term performance. In this way, employees are more likely to be motivated and release the innovation potential, and thus promote the innovation of MSEs. The results confirmed H2a.

\subsubsection{Strategic Investment to MSEs human capital}

Human capital is another important factor affecting the innovation of MSEs. The development of fintech can help MSEs increase human capital investment and promote the upgrading of enterprise human capital, thus promoting enterprise R\&D and innovation. The recruitment of higher-skilled and creative thinking employees such as university graduate students are important to MSEs innovation for their distance to the human capital frontier is short. When the incentive of professional technicians and the staff training enhances, enterprises will gain strong innovation power. Table 9 reports the estimation results. The dependent variables in the column (1)-(3) are the number of university graduate students recruitment, the income of professional technicians in the enterprise and the employee training fee that the enterprise provides, respectively. From the estimation results, it can be seen that fintech significantly promotes the university graduate students recruitment, and increases the technicians income and the training fee. These results are significant at the 1\%–10\% statistical level. Digital finance alleviates the financing constraints and increases capital as well as total wages of MSEs which makes it possible for MSEs to hire more employees who are higher-skilled and creative thinking (representing by university graduate students) and increase the human capital investment. Through upgrading the enterprise human capital, MSEs can obtain high-quality labor which is necessary for R\&D and innovation, thus promoting enterprises to engage in R\&D and innovation activities. The results confirmed H2b.

\subsection{Heterogeneity of the Impact of Fintech on Innovation of MSEs}

In order to further investigate which type of MSEs are more affected by fintech, this paper will then divide the sub-samples according to the location, the age, the assets and the size of the employees of the MSEs to conduct a more detailed heterogeneity test. Table 10 reports the estimation results of the impact of fintech development on the innovation heterogeneity of MSEs. Heterogeneity effects in different regions are reported in columns (1)–(3) of Table 10. It can be seen from the estimation results that the impact of fintech on MSEs located in eastern region (developed region) of China is significantly positive, while the impact on the MSEs which located in central and western regions (developing regions) of China are not significant. This indicates that fintech still plays a significant role in promoting the innovation of MSEs in the eastern region under the condition that other factors remain unchanged. The results support H3a.

We will then group according to the age of enterprises to examine the impact of fintech on MSEs of different ages. Specifically, according to the 50th percentile of the age of enterprises, that is, the survey year (2015) minus the year of enterprises registration equals 5 years, the sample is divided into the sample of enterprises aged 5 years or younger and more than 5 years. It can be seen from the estimation results in columns (4)–(5) in Table 10 that fintech has a greater impact on MSEs with an enterprise age of 5 years or younger. Specifically, for every increase of 100 units in the fintech index, the probability of MSEs aged 5 years and younger engaged in R\&D and innovation activities will increase by 46\%, which is 24.7 percentage points higher than the 21.3\% of MSEs over 5 years. It can be seen that fintech has a greater impact on the innovation of MSEs with shorter established years, which further confirms the inclusive impact of fintech. The results support H3b.

From the perspective of enterprise size, we divide samples below and above 50th percentile into small and large enterprises according to the size of assets and employees. As can be seen from columns (6)–(7) of Table 10, the development of fintech has a significant positive impact on the two groups of enterprises with different asset sizes, and the significance level is 1\%. Fintech, by contrast, has a bigger impact on enterprises with fewer assets. Specifically, for every 100 units increase in the fintech index, the probability of enterprises with fewer assets engaging in R\&D and innovation activities will increase by 40.8\%, 11 percentage points higher than that of larger assets enterprises. As can be seen from columns (8)–(9), the development of fintech also has a significant positive impact of 1\% on two groups of enterprises with different employee sizes. In contrast, fintech has a greater impact on enterprises with small size of employees. Specifically, for every 100 units increase in the fintech index, the probability of enterprises with small size of employees engaging in R\&D and innovation activities will increase by 35.3\%, which is 2.5 percentage points higher than the 32.8\% of enterprises with large size of employees. The results support H3c and H3d. It can be seen that the development of fintech has a greater impact on the innovation of MSEs with relatively small assets and employees, which further confirms the inclusion of the development of fintech, either.

\subsection{Further Analysis}

In order to further understand the relationship between fintech and innovation of MSEs, this paper further analyzes the impact of fintech on different forms of R\&D and innovation activities—independent R\&D, commissioned R\&D, technology introduction and cooperative R\&D. The estimation results are shown in Table 11. It can be seen from the estimated coefficient in the column (1) that every 100 units increase of the fintech index will increase the probability of enterprises engaging in independent research and development activities by 20.1\%, and the impact is significant at the level of 5\%. It can be seen from the estimation coefficient in the column (3) that the probability of technology introduction of enterprises decreases by 14.7\% for every 100 units increase in the fintech index, and the influence is significant at the level of 10\%. Column (2) and (4) show that the development of fintech has no significant impact on the commissioned R\&D and cooperative R\&D of enterprises. The development of fintech significantly promotes the probability of MSEs engaging in independent R\&D and innovation activities, and reduces their dependence on technology introduction. This is of great practical significance to cultivate independent innovation ability and enhance the core competitiveness of MSEs in developing countries.

\section{Implications}

\subsection{Theoretical implications}

Different from the existing literature, the theoretical contributions of this study are as follows. Firstly, this paper expands the research on financial development and enterprise innovation from the perspective of fintech. Some of previous studies have examined the effect of traditional finance on enterprise innovation. This paper examines the impact of the integration of emerging technology and traditional finance on enterprise innovation from the perspective of fintech, which is an important supplement to previous studies.

Secondly, this study makes an important supplement for the research of fintech at the micro-economic level. Contrary to the literature that focus on the impact at the macro level, this paper investigates the impact of fintech on MSEs at the micro level, which enriches the empirical research on the impact of fintech at the micro level.

Thirdly, this paper verifies the inclusive impact of fintech on MSEs. Different from the literature that studies the impact of fintech on the listed companies, this paper focuses on MSEs and uses the data of large national MSEs survey to answer the question of whether fintech is inclusive, which is an important supplement to the research on the impact of fintech.

Lastly, this paper systematically investigates the influence mechanism of fintech in promoting innovation of MSEs, and finds that fintech promotes MSEs innovation through two channels: alleviating the financing constraints and upgrading the human capital of MSEs. Specifically, digital finance raises the proportion and income of professional technicians, and promotes the investment in staff training in MSEs, so as to promotes the innovation of MSEs, especially the innovation of independent R\&D.

\subsection{Practical implications}

This paper also has important policy implications. MSEs play an important role in most economies, particularly in developing countries. MSEs not only contribute a lot of innovation to the country, but also provide a lot of employment opportunities, which is an inexhaustible driving force of global economic growth. However, the restrictions of traditional financing channels are the main obstacle for MSEs. The emergence and rapid development of fintech provide the important way to solve the financing problems of MSEs. Compared with traditional finance, fintech has better geographical penetration and stronger user accessibility, which can effectively reduce the financing costs of MSEs and expand the coverage breadth of financial services. Therefore, the government should further broaden the financing channels of small and micro enterprises, expand the coverage and depth of fintech, unlock sources of capital, and provide inclusive financial services for MSEs. Especially under the impact of COVID-19, MSEs should be released to innovate and stimulate the market economy, so as to provide a source of vitality for global economic recovery and high-quality development.

\subsection{Limitations and Prospects of the Future Research}

There are some limitations of this study. The database of CMES used in this paper is currently only one-year data (2015) and we can only use the cross-section data of MSEs for this study. The panel data with time trend of fintech development if available may be able to fulfil more complete studies. Hence, there are some issues related to time trends and the dynamic effects which need to be further studied and we look forward to the higher quality panel data for MSEs in the future. The location of the MSEs, as private information of enterprises, should be protected during the data survey. So only the province where the enterprise is located is disclosed in CMES database. The paper match the information at the provincial level. In the future if there the desensitized location information can be disclosed, more detailed analysis can be conducted.

\section{Conclusions}

This paper gives empirical evidence on the impact of fintech on the micro and small enterprises (MSEs) using the Peking University Digital Financial Inclusion Index of China (PKU-DFIIC) and the database of China Micro and Small Enterprise Survey (CMES). We find that the development of fintech significantly promotes the innovation of MSEs as well as increases the innovation input or output of MSEs. The results are stable and reliable by using the instrumental variables (objective geographical distance from the enterprise location to coastal port and the internet penetration rate) to deal with the endogeneity problems.

The underlying mechanisms lie in that fintech development promote the long-term strategic incentives and investment to human capital, such as choosing promotion and stock as the incentives of employees and recruiting university graduate students. In the heterogeneity analysis, we find that fintech has a greater impact on those MSEs which are located in developed regions, aged 5 years or younger, and with fewer assets and employees. Further analysis shows that fintech promotes the innovation of MSEs independent research and development (R\&D). By examining the influence of fintech on different forms of R\&D and innovation activities, it is found that fintech significantly improves the probability of independent innovation and reduces the dependence of enterprises on the introduction of technology.

Based on existing research on the impact of fintech on enterprise innovation, we verify the inclusion impact of fintech on the MSEs, and further clarifies the influence mechanism of fintech to promote enterprise innovation. Under the background of rapid development of fintech and the impact of COVID-19, the research provides a policy basis for strengthening inclusive financial support for MSEs and activating innovation power of MSEs.

\bibliographystyle{unsrt}


\end{document}